# Ehancement of optical forces at bound state in the continuum

Haoye Qin,[1] Walid Redjem,[1] Boubacar Kante[1, *]

[1] *Department of Electrical Engineering and Computer Sciences, University of California, Berkeley, California 94720, United States*
*Corresponding author:* bkante@berkeley.edu

**Abstract:**

Light-actuated motors, vehicles, and even space sails have drawn tremendous attention for basic science and applications in space, biomedical, and sensing domains. Optical bound states in the continuum (BIC) are topological singularities of the scattering matrix, known for their unique light trapping capability and enhanced light-matter interaction. We show that BIC modes enable the generation of enhanced and tunable optical forces and torques. A sharp and controllable lineshape is observed in forces and torques spectra when approaching high-Q resonance BIC modes. Wavelength and polarization tunability are presented as an effective method to control the forces on BIC enclosed structures. Finally Finite-size simulations are performed to evaluate the practical application for a BIC assisted metavehicle.

Bound states in the continuum (BIC) have attracted considerable interest due to their ability to trap waves within the continuum without radiation loss [1]. First proposed in quantum mechanics, bound state in the continuum have naturally been extended to optics [4-6]. Especially, periodic optical platforms such as metasurfaces, gratings, and photonic crystal have enable practical demonstrations and applications of BICs, with high Q-factor and large mode volume [1–3]. For instance, BICs have been used for directional and single mode lasing operation [7,8], biosensing [9] and enhancing non-linear effects [11-13].

Light can exert forces and torques as illustrated in optical tweezers [14] and particle-trapping devices [15,16]. Metasurfaces provide unique opportunities to engineer and manipulate light [17], thus leading to the conception of light-based propulsion for space travel, i.e., laser and solar-driven spacecraft. Recently, novel studies on optical force contributed to the proposal of metavehicles [18], plasmonic motors [19–21], self-stabilizing photonic structure for levitation and propulsion [22], and lightsails [23–25], which rely on asymetric scattering of light [22]. However, metasurfaces showing large tunability and enhancement of optical forces and torques have been missing while such demonstration could be important in building up effective light vehicles/motors/sails with full controllability via properties of incident light.

In this paper, the strong field localization, polarization singularity, and high Q-factor induced by the BIC resonance are employed to enhance and tune optical forces and torques, which can directly benefit the development of optically driven transportation. We introduced an asymmetry between two coupled dielectric particles to form a high-Q BIC mode. By controlling the wavelength and/or polarization of the incident light, the generated force and torque can switch direction and intensity, which is useful for on-demand all-optical manipulation. Finally, a realistic finite-size simulation is conducted to quantify the BIC enhanced and tunable optical forces to control a metavehicle.

The designed metasurface unit-cell for BIC mode is shown in Fig. 1(a). Two silicon bars with the same width (200 nm) and length $L_1$ (fixed at 280 nm) and $L_2$ are placed in free space. The gap between the two bars is 75 nm and their thickness is 250 nm. Periodic boundary conditions are applied in x and y direction. $L_2$ can be tuned to change the asymmetry ratio defined as $\alpha = (L_1 - L_2)/L_2$ [26]. When $\alpha = 0$, in-plane symmetry is fulfilled, and a BIC mode is present. Reducing the length of the one bar breaks the in-plane symmetry and leads to a quasi-BIC mode with a sharp spectral resonance. To investigate the optical force and torque associated with the high-Q quasi-BIC mode, we integrated the Maxwell stress tensor (MST) around a box enclosing the unit cell [Fig. 1 (a)]. The expressions used for calculating the time-averaged force **F** and torque **Tor** are given by:

$$\mathbf{F} = \iint_S (\mathbb{T} \cdot \boldsymbol{n}) \mathrm{dS}, \tag{1}$$

$$\mathbf{Tor} = \iint_S \boldsymbol{r} \times (\mathbb{T} \cdot \boldsymbol{n}) \mathrm{dS} \tag{2}$$

where S is the surface of the MST box, $\boldsymbol{n}$ is the unit vector normal to the surface, $\boldsymbol{r}$ is the position vector regarding the original point. $\mathbb{T}$ is the stress tensor with matrix elements defined as:

$$\mathbb{T}_{ij} = \epsilon_0 \left( E_i E_j - \frac{1}{2} |E|^2 \delta_{ij} \right) + \frac{1}{\mu_0} \left( B_i B_j - \frac{1}{2} |B|^2 \delta_{ij} \right), \tag{3}$$

where $E_i$ and $B_i$ are the $i^{\text{th}}$ component of the electric and magnetic field respectively. In the following all numerical electromagnetic simulation are performed using the commercial finite-element-method solver COMSOL Multiphysics.

Fig. 1(b-d) illustrate the optical response of the metasurface with two asymmetric particles as a unit-cell. In Fig. 1(b), the evolution of transmission spectrum is shown for different asymmetry ratio. As expected, at $\alpha = 0$, in-plane symmetry produces a symmetry-protected BIC mode with disappeared linewidth. Increasing $\alpha$ leads first to a sharp resonance dip with decreasing quality factor illustrated by the broadened and blue-shifted resonance. Green (individual mode 1) and blue (individual mode 2) markers in Fig. 1 (b) indicate the wavelength evolution of two individual modes versus the change of asymmetry ratio. Fig. 1(d) shows the exponentially decreasing Q-factor of the collective mode extracted using a rational fit method for the complex scattering parameter $S_{21}$. An interesting phenomenon accompanied with the "visible" collaborative quasi-BIC mode is that two individual quasi-BIC modes have emerged. The two modes can be seen in the enlarged view of the spectrum [Fig. 1(c)], featuring sharp lineshape They arise from the in-plane symmetry of individual particle instead of the two interactive particles and are then perturbed by the mode coupling to reveal quasi-BIC feature. The individual mode 2 is more sensitive to the asymmetry ratio because it corresponds to the resonator that is tuned in dimension. Corresponding magnetic field profiles for individual mode 1 and 2 plotted in Fig. 1(c) confirm the mode localization in only one particle. The near-field distribution of the magnetic field for the collaborative quasi-BIC mode is presented in the top inset of the figure and the strongly confined field profile indicates a high-Q resonance. It is named a collaborative mode as there is a clear interaction between the two resonant particles.

In Fig. 2 (a,b), a strong enhancement of optical forces $F_x$ and $F_y$ assisted by the high-Q resonance of the quasi-BIC mode is shown when the structure has a small asymmetry. The force is calculated under an incident power of 1 µW/µm². For $\alpha = 0.05$ the optical force is ten times larger than for $\alpha = 0.20$. As the asymmetry ratio increases, the optical force reduces exponentially. In addition, tunability can be achieved by sweeping the incident light wavelength. There exists a sharp peak switch from positive to negative $F_y$ and crossing zero exactly at the BIC resonance. This is better seen in Fig. 2(d) showing the optical force $F_y$ around the BIC resonance (1055 nm). The positive to negative feedback region is due to the asymmetry of the field along the $xz$ plane that induces a force along y [see Eq. (3)]. At the BIC wavelength the field is symmetric thus $F_y$ is null. Tunability of optical forces and torques can also be achieved through varying polarization of the incident light. Fig. 2(c,d) demonstrate the effect of polarization on optical forces for a given asymmetry ratio of 0.05. Changing the polarization from $0.1\pi$ to an angle of $0.3\pi$ and y polarization results in increasing the BIC induced force. Fig. 2(e,f) present the tailorable optical torque of the metasurface calculated from Eq. (2) under illumination of different polarization states. The direction of the torque can be tuned with the wavelength and the polarization angle, enabling a full control of the structure's rotation and angular position. The effect of circular polarized light (LCP, RCP) is evaluated in Fig. 2(f). A switch from LCP to RCP also symmetrically switches the direction of the torque, hence the direction of rotation.

A practical design with $20 \times 20$ unit-cells on glass substrate is now investigated. The finite size structure comprises a bottom layer of glass 0.5 µm thick and previously studied asymmetric particles with asymmetry ratio of 0.09. Simulated transmission of Fig. 3(a) shows a quasi-BIC resonance dip corresponding to a Q-factor of 1135, which is comparable with the Q-factor of 1528 previously extracted from an asymmetric infinite structure [Fig. 1(d)]. This relatively small reduction is due to additional scattering losses at the boundaries of the finite structure. The top view and side view of this quasi-BIC mode are presented in Fig. 3(c,d) with electrical field distribution. The mode is highly localized within the central nanobars and along one direction due to the asymmetry of the unit-cells. We can observe scattering mainly at the right edges of the structure where the field is stronger, which explain the reduction of Q-factor mentioned above. Fig. 3(b) shows the enhancement and tunability of optical forces and torque. The intensity of the forces is comparable to the case of the infinite periodic metasurface.

To investigate the metasurface as an optical motor in water, we assume that only the friction coming from the bottom surface of the metavehicle is present and we neglect the frontal area friction because of its large length-thickness ratio (20:1). Thus, under the Laminar flow approximation, the total drag coefficient can be estimated from the local skin coefficient:

$$C_D = \frac{1}{bL} \int_0^L 0.664 \left(\frac{U}{\nu}\right)^{-0.5} x^{-0.5} b dx = 1.33 \left(\frac{\nu}{UL}\right)^{0.5} \quad (4)$$

or,

$$C_D = \frac{1.33}{\sqrt{Re_L}} \quad (5)$$

where $Re_L = \frac{UL}{\nu}$ is called Reynolds number, and U, L, b, ν represents the flow velocity, length, depth of the metasurface, and kinematic viscosity of the flow, respectively [27]. For water at room temperature, $\nu = 8.917 \times 10^{-7} m^2/s$ resulting in $Re_L = 16.26U$ and $C_D = \frac{0.3798}{\sqrt{U}}$ for propulsion in y direction. Using Eq. (5) for the drag coefficient and assuming a balanced force at final state, the achieved velocity with optical propulsion in y direction and with an incident power of 1 µW/µm² is evaluated [Fig. 4(b)] for different loaded mass ratio – corresponding to the ratio between the loaded mass and the mass of the metavehicle. Due to the high Q-factor and large force enhancement, our proposed device can move relatively fast even with a loaded mass comparable or larger than the mass of the vehicle. Fig. 4(a) shows the schematic of the metasurface with BIC enhanced and tunable optical forces applied to the metavehicle for transportation and delivery. An object is loaded on the metavehicle to illustrate its carrying capacity [Fig.4 (b)]. Finally, Fig. 4(c) presents the tunability of the metavehicle for controlling the direction of motion and rotation via wavelength, polarization angle, and polarization state as illustrated in Fig. 2 and Fig. 3.

We have presented the incorporation of quasi-BIC mode into metasurfaces for enhancing the optical forces and torques with large tunability. By changing the asymmetry of the two-bar structure, high-Q quasi-BIC mode can be revealed showing enhanced force in x, y direction and torque in z direction. Along with this enhancement, wavelength and polarization tuning of force/torque directions can be realized. Finally, the finite size simulation showed that even a relatively small structure can be an efficient metavehicle. Hence, the proposed metasurface may contribute to on-demand control of light vehicles/motors/sails for applications in space exploration, medical, sensing, and actuators by only modulating properties of incident light.

**References**

1. C. W. Hsu, B. Zhen, A. D. Stone, J. D. Joannopoulos, and M. Soljacic, Nat. Rev. Mater. **1**, (2016).
2. K. Koshelev, A. Bogdanov, and Y. Kivshar, Sci. Bull. **64**, 836 (2019).
3. D. C. Marinica, A. G. Borisov, and S. V. Shabanov, Phys. Rev. Lett. **100**, 1 (2008).
4. Y. Plotnik, O. Peleg, F. Dreisow, M. Heinrich, S. Nolte, A. Szameit, and M. Segev, Phys. Rev. Lett. **107**, 28 (2011).
5. C. W. Hsu, B. Zhen, J. Lee, S. L. Chua, S. G. Johnson, J. D. Joannopoulos, and M. Soljačić, Nature **499**, 188 (2013).
6. Thomas Lepetit and Boubacar Kante, Phys. Rev B, 90, 241103
7. A. Kodigala, T. Lepetit, Q. Gu, B. Bahari, Y. Fainman, and B. Kanté, Nature **541**, 196 (2017).
8. C. Huang, C. Zhang, S. Xiao, Y. Wang, Y. Fan, Y. Liu, N. Zhang, G. Qu, H. Ji, J. Han, L. Ge, Y. Kivshar, and Q. Song, Science (80-. ). **367**, 1018 (2020).
9. Ndao, Abdoulaye, Hsu, Liyi, Cai, Wei, Ha, Jeongho, Park, Junhee, Contractor, Rushin, Lo, Yuhwa and Kanté, Boubacar. *Nanophotonics*, vol. 9, no. 5, 2020, pp. 1081-1086.
10. M. F. Limonov, M. V. Rybin, A. N. Poddubny, and Y. S. Kivshar, Nat. Photonics **11**, 543 (2017).
11. L. Carletti, K. Koshelev, C. De Angelis, and Y. Kivshar, Phys. Rev. Lett. **121**, 33903 (2018).
12. N. Bernhardt, K. Koshelev, S. J. U. White, K. W. C. Meng, J. E. Fröch, S. Kim, T. T. Tran, D. Y. Choi, Y. Kivshar, and A. S. Solntsev, Nano Lett. **20**, 5309 (2020).
13. K. Koshelev, Y. Tang, K. Li, D. Y. Choi, G. Li, and Y. Kivshar, ACS Photonics **6**, 1639 (2019).
14. K. B. Crozier, Light Sci. Appl. **8**, 4 (2019).
15. J. C. Ndukaife, A. V. Kildishev, A. G. A. Nnanna, V. M. Shalaev, S. T. Wereley, and A. Boltasseva, Nat. Nanotechnol. **11**, 53 (2016).
16. D. Gao, W. Ding, M. Nieto-Vesperinas, X. Ding, M. Rahman, T. Zhang, C. T. Lim, and C. W. Qiu, Light Sci. Appl. **6**, (2017).
17. D. Neshev and I. Aharonovich, Light Sci. Appl. **7**, 1 (2018).
18. D. G. Baranov, S. Jones, G. Volpe, and R. Verre, 1 (n.d.).
19. O. Ilic, I. Kaminer, B. Zhen, O. D. Miller, H. Buljan, and M. Soljačić, Sci. Adv. **3**, 1 (2017).
20. Z. Zhan, F. Wei, J. Zheng, W. Yang, J. Luo, and L. Yao, Nanotechnol. Rev. **7**, 555 (2018).
21. M. Liu, T. Zentgraf, Y. Liu, G. Bartal, and X. Zhang, Nat. Nanotechnol. **5**, 570 (2010).
22. O. Ilic and H. A. Atwater, Nat. Photonics **13**, 289 (2019).
23. W. Jin, W. Li, M. Orenstein, and S. Fan, ACS Photonics **7**, 2350 (2020).
24. K. V. Myilswamy, A. Krishnan, and M. L. Povinelli, Opt. Express **28**, 8223 (2020).
25. H. A. Atwater, A. R. Davoyan, O. Ilic, D. Jariwala, M. C. Sherrott, C. M. Went, W. S. Whitney, and J. Wong, Nat. Mater. **17**, 861 (2018).

26. K. Koshelev, S. Lepeshov, M. Liu, A. Bogdanov and Y. Kivshar, Phys. Rev. Let. 121, 193903 (2018)
27. F. M. White, 7th ed. (McGraw Hill, 2011).

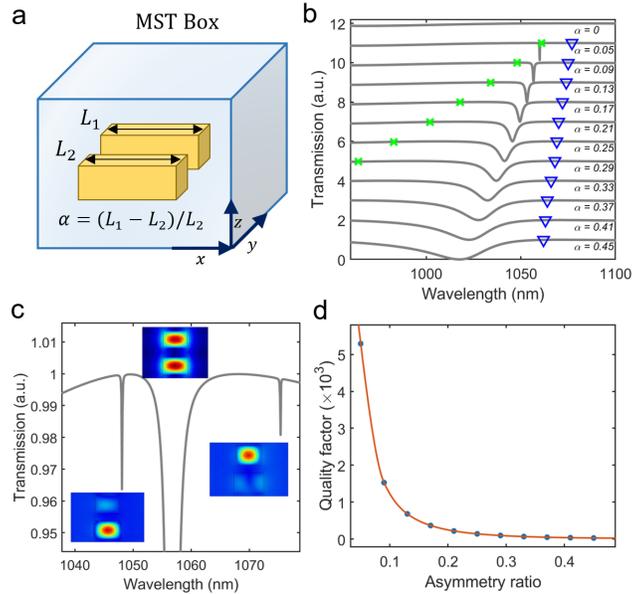

**Fig. 1.** Schematic and optical response of the coupled particles exhibiting a BIC mode for enhanced optical force and torque. (a) Schematic view of the asymmetric silicon bars with length $L_1$ (fixed at 280 nm) and $L_2$. Both bars have a width of 200 nm. The gap between them is 75 nm and their common thickness is 250 nm. The asymmetry ratio $\alpha$ is defined as $\alpha = (L_1 - L_2)/L_1$. Periodic boundary condition is applied in x and y direction with a lattice constant $p_x = 200$ nm and $p_y = 530$ nm respectively. (b) Transmission spectrum for different asymmetry ratio. The dip corresponds to the quasi-BIC mode whereas the green cross marker and blue triangle corresponds to individual modes 1 and 2 shown in (c). (c) Enlarged view of the two individual BIC modes with $\alpha = 0.09$ in (b). Corresponding distribution of magnetic field for individual BIC modes 1 and 2 and for the collective BIC mode is shown above, respectively. For individual modes field is localized in either the top or bottom particle. (d) Quality factor as a function of asymmetry ratio for the collective mode. The blue dots are the extracted quality factor through rational fitting. The red line is an exponential fit.

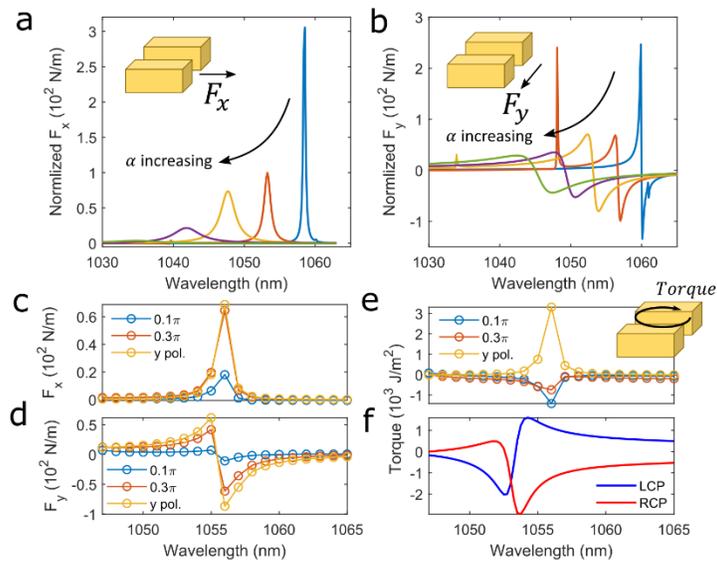

**Fig. 2.** Optical force in x (a) and y (b) direction as a function of wavelength for different asymmetry ratios [0.05 (blue), 0.09 (red), 0.13 (yellow), 0.17 (purple) and 0.21 (green)]. The forces calculated from MST box are normalized to the period of the unit cell. Inset shows the generated optical force in x direction. Optical force $F_x$ (c) and $F_y$ (d) with α = 0.13 for different polarization of the incident light. The optical force can be switched from positive to negative via tuning the incident light wavelength and force intensity via polarization angle. (e,f) Optical torque for different polarization states. (e) linear polarization with different polarization angles. (f) evaluation of optical torque under left and right circular incident polarized light.

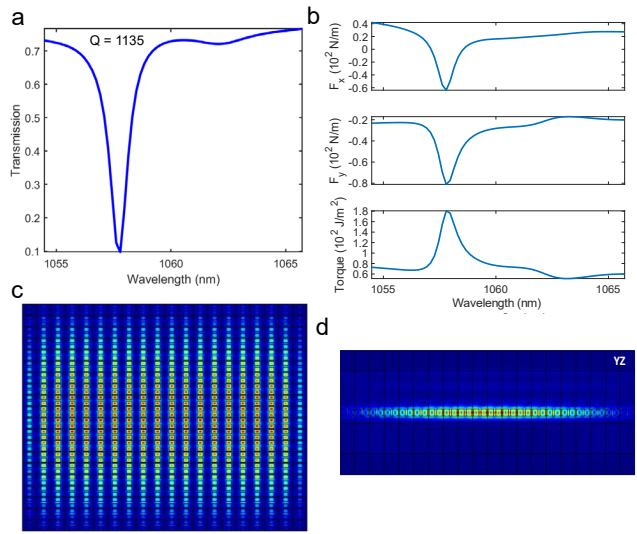

**Fig. 3.** Proposed meta-motor on glass substrate. (a) Transmission spectrum from finite-size simulation showing a Q factor of 1135. (b) Calculated optical force and torque from finite-size simulation. (c) Top view and (d) side view show the electric filed distribution of the corresponding quasi-BIC mode present in the transmission spectra.

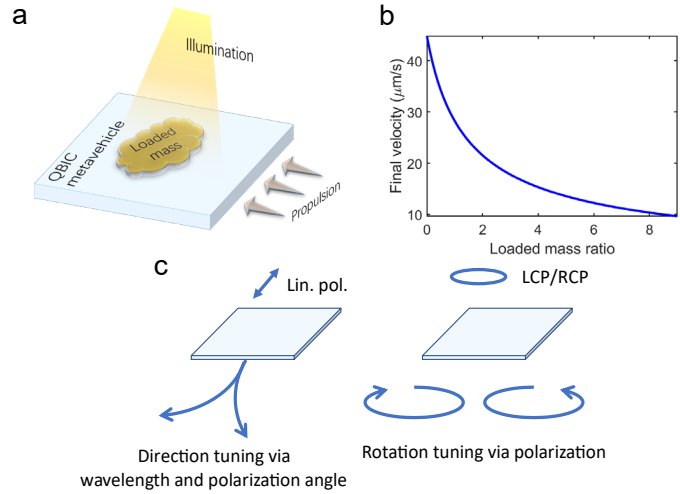

**Fig. 4.** (a) Schematic of the BIC assisted metavehicle with a loaded mass for usage of transportation and delivery. (b) Estimated final-state velocity for propulsion in y direction as a relation of additional loaded mass with an incident power of $1~\mu W/\mu m^2$. The loaded mass is given in term of unloaded mass of the metavehicle. (c) Tunability of the BIC assisted metavehicle for controlling moving direction and rotation via wavelength, polarization angle, and polarization state

}